\documentstyle[prl,multicol,aps,epsf]{revtex}

\newcommand{\NCCO}{$\bf Nd_{1.85}Ce_{0.15}CuO_{4-y}$\ }
\newcommand{\ncco}{Nd$_{1.85}$Ce$_{0.15}$CuO$_{4-y}$}

\newcommand{\srtio}{SrTiO$_3$\ }

\begin{document}

\draft
\sf

\title{Superconducting \NCCO Bicrystal Grain Boundary Josephson Junctions}

\author{\renewcommand{\thefootnote}{\alph{footnote})}
S.~Kleefisch, L.~Alff\footnote{e-mail: alff@ph2.uni-koeln.de}, U.~Schoop,
A.~Marx, and R.~Gross}

\address{II.~Physikalisches Institut, Universit\"{a}t zu K\"{o}ln,
Z\"{u}lpicherstr.~77, D - 50937 K\"{o}ln, Germany}

\author{M.~Naito and H.~Sato}

\address{NTT Basic Research Laboratories, 3-1 Morinosato Wakamiya,
Atsugi-shi, Kanagawa 243, Japan}

\date{received January 15, 1998}

\maketitle

\begin{abstract}

We have studied the electric transport properties of symmetrical [001] tilt
\ncco\ (NCCO) bicrystal grain boundary Josephson junctions (GBJs) fabricated
on SrTiO$_3$ bicrystal substrates with misorientation angles of 24$^{\circ}$
and $36.8^{\circ}$.  The superconducting properties of the NCCO-GBJs are
similar to those of GBJs fabricated from the hole doped high temperature
superconductors (HTS).  The critical current density $J_c$ decreases
strongly with increasing misorientation angle.  The products of the critical
current $I_c$ and the normal resistance $R_n$ ($\sim 100\,\mu$V at 4.2\,K) are
small compared to the gap voltage and fit well to the universal scaling law
$I_cR_n \propto \sqrt{J_c}$ found for GBJs fabricated from the hole doped HTS.
This suggests that the symmetry of the order parameter, which most likely is
different for the electron and the hole doped HTS has little influence on the
characteristic properties of symmetrical [001] tilt GBJs.

\end{abstract}

\pacs{PACS: 74.25.Fy, 74.50.+r}

\vspace*{-9.5cm}\noindent
Applied Physics Letters \hfill to be published June 1, 1998
\vspace*{8.5cm}

\begin{multicols}{2}
\narrowtext

Bicrystal grain boundary Josephson junctions (GBJs) have been studied
intensively using epitaxial thin films of the various hole doped high
temperature superconductors (HTS) \cite{Dimos:90,Gross:95,Beck:96}. However,
there is very limited information on GBJs fabricated from the electron doped
material \ncco (NCCO)\cite{Kawasaki:93}. For the vast majority of GBJs the
transport properties can be well described within the intrinsically shunted
junction (ISJ) model\cite{Gross:95,Gross:91a,Marx:97}.  In this model a
continuous, but spatially inhomogeneous insulating grain boundary barrier is
assumed that contains a high density of localized defect states.  The
localized states allow for resonant tunneling of quasiparticles providing an
intrinsic resistive shunt, whereas resonant tunneling of Cooper pairs is
prevented by Coulomb repulsion.  This results in reduced $I_cR_n$ products and
a scaling behavior $I_cR_n \propto \sqrt{J_c}$ observed in many experiments
\cite{Gross:95,Gross:91a,Marx:97,Gross:90a}.  Meanwhile, it is well
established that most hole doped HTS have a dominant $d$-wave component of the
order parameter (OP) \cite{vHarlingen:95}.  The influence of the $d$-wave
symmetry of the OP on the magnetic field dependence of $I_c$ has been shown
for asymmetric $45^{\circ}$ tilt YBCO-GBJs \cite{Mannhart:96}.  So far, in the
ISJ-model the likely $d$-wave pairing state of the hole doped HTS has not been
taken into account and the relevance of the $d$-wave symmetry of the OP for
the characteristic properties of symmetrical [001] tilt GBJs such as the small
value and the scaling behavior of the $I_cR_n$ product is still a point of
controversy.  To clarify this issue we have studied [001] tilt NCCO-GBJs.
Since for the electron doped material NCCO there is convincing experimental
evidence for a $s$-wave symmetry of the OP
\cite{Huang:90,Wu:93,Andreone:94,Alff:97}, NCCO-GBJs represent an interesting
model system to test the influence of the OP symmetry on the transport
properties of GBJs.

The NCCO-GBJs were fabricated by molecular beam epitaxy (MBE) of $c$-axis
oriented NCCO thin films on \srtio bicrystal substrates.  The substrate
temperature during growth was about 730$^{\circ}$C and ozone was used as
oxidation gas.  For the critical temperature $T_c$, the resistivity
$\rho$(250\,K), and $\rho$(25\,K) typical values of 23 - 24\,K,
$350\,\mu\Omega$cm, and $50\,\mu\Omega$cm were obtained, respectively.  A
detailed description of the fabrication process was given by Naito {\em et
al.} \cite{Naito:95,Yamamoto:97}.  For NCCO-GBJs the film quality is a key
issue, since small deviations from optimum (preferentially oxygen
stoichiometry) can reduce $I_c$ to unmeasurably small values.  More precisely,
for small $I_c$ the Josephson coupling energy $E_J$ becomes comparable to or
even smaller than the thermal energy $k_BT$.  This may be the reason why until
now a finite $I_c$ has been observed only for low angle ($\le 10^{\circ}$)
NCCO-GBJs\cite{Kawasaki:93}.  For our samples the direct observation of a
critical current was possible only for the $24^{\circ}$ tilt GBJs having $J_c$
values between 5 and 50\,A/cm$^2$ at 4.2\,K. For the $36.8^{\circ}$ tilt GBJs
the $J_c$ values are more than an order of magnitude smaller preventing the
direct measurement of $I_c$ due to thermal noise effects ($E_J<k_BT$).
Although we cannot give any functional dependence at present, our data clearly
show that $J_c$ strongly decreases with increasing misorientation angle.

\begin{figure}[bt]
\noindent\hspace*{0cm}
\centering\epsfxsize=9cm\epsffile{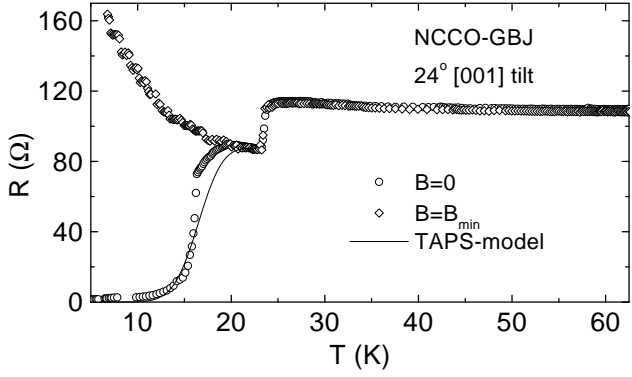}\vspace*{-1.0cm}
\caption[]{\small Resistance vs.~temperature curve of a symmetrical 24$^{\circ}$ [001]
tilt NCCO bicrystal GBJ ($W$=10\ $\mu$m) for $B=0$ and $B=B_{min}$. The solid
line shows the prediction of the RSJ-model including thermal noise
\cite{Ambegaokar:69,Gross:90}. }
\label{fig:rtfoot}
\vspace*{-0.5cm}
\end{figure}

Fig.~\ref{fig:rtfoot} shows a typical resistive transition of a NCCO-GBJ. The
onset temperature of 24\,K marks the resistive transition of the NCCO film.
The foot structure emerging below about $90\,\Omega$ is close to resistance
$R_p(T)=R_n(T)\left[{\rm I_0}\left(\frac{\hbar I_c(T)}{2ek_BT}\right)
\right]^{-2}$, expected for ideal overdamped Josephson junctions due to
thermally activated phase slippage (TAPS) \cite{Ambegaokar:69,Gross:90}. Here,
${\rm I_0}$ is the modified Bessel function of the first kind.  As shown by
Fig.~\ref{fig:rtfoot}, the TAPS model qualitatively describes the measured
foot structure using $R_n(T)=const.=90\,\Omega$ and $I_c(T)=
I_c(0)(1-T/T_c)^2$.  Certainly, a better fit is possible by taking into
account a temperature dependent $R_n$ and a modified $I_c(T)$ dependence. The
normal resistance times junction area, $\rho _n$, of the $24^{\circ}$ tilt
GBJs ranged between 1.5 and $8\times10^{-6}\,\Omega$cm$^2$ and was more than
an order of magnitude larger for the $36.8^{\circ}$ tilt GBJs.


\vspace*{-0.2cm}



\vspace*{-0.2cm}
\begin{figure}[tb]
\noindent\hspace*{0cm}
\centering\epsfxsize=9cm\epsffile{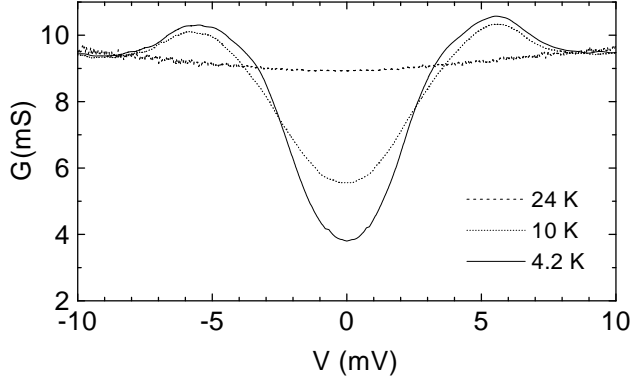}\vspace*{-1.0cm}
\caption[]{\small Conductance versus voltage curves of a 24$^{\circ}$ [001] tilt NCCO
bicrystal GBJ at different temperatures.
}
\label{fig:gv}
\vspace*{-0.4cm}
\end{figure}


Fig.~\ref{fig:gv} shows the $G(V)=dI(V)/dV$ curves of a NCCO-GBJ.  The
Josephson current was suppressed by applying a small magnetic field $B_{min}$
corresponding to a minimum of the $I_c(B)$ dependence of the GBJ.  A clear gap
structure is observed in the $G(V)$ curves with a gap voltage $V_g=\Delta
_0/e$ \cite{Gapvoltage} of about 6\,mV that is comparable to values reported
in literature \cite{Wu:93,Andreone:94,Alff:96c,Naito:98}.  The observation of
a clear gap structure and the temperature independent conductance well above
$V_g$ strongly suggest that the dominating transport mechanism in the
NCCO-GBJs is elastic tunneling.  That is, in agreement to what is observed for
YBCO-, LSCO- and BSCCO-GBJs \cite{Gross:95,Froehlich:97a}, also for the
NCCO-GBJs the dominant transport mechanism most likely is elastic tunneling
via a single localized state.  Assuming purely elastic resonant tunneling a
density of localized states $n_{res}=h/2e^2\rho_n\approx 1\times
10^{14}$\,m$^{-2}$ can be estimated which is close to values reported
recently\cite{Froehlich:97a}. We also note that in contrast to GBJs fabricated
from the hole doped HTS, the $G(V)$ curves of the NCCO-GBJs show no zero bias
conductance peak which is consistent with a $s$-wave symmetry of the OP in
NCCO \cite{Alff:96c}.  We note, however, that the OP of NCCO may be highly
anisotropic, i.e., the amplitude of the OP may show a strong $k$-dependence
but there is no sign change as for a $d$-wave OP.

Suppressing $I_c$ by applying $B=B_{min}$ we have $\hbar I_c(B)/2ek_BT \ll 1$
and, hence, $R_p(T) \simeq R_n(T)$, i.e., $R_n(T)$ can be measured directly.
The result is shown in Fig.~\ref{fig:rtfoot} for a $24^{\circ}$ tilt GBJ. A
very similar result is obtained for the $36.8^{\circ}$ tilt GBJs even for
$B=0$, since here $\hbar I_c(B)/2ek_BT \ll 1$ even at zero field.  It is
evident that $R_n$ increases strongly with decreasing temperature.  We
emphasize that this increase of $R_n$ cannot be caused by the presence of
inelastic tunneling processes freezing out on going to lower temperatures,
since the dominating transport mechanism is elastic tunneling as discussed
above.  In order to discuss the origin of the observed $R_n(T)$ dependence we
have to take into acount that $R_n$ corresponds to $dV/dI$ at $V\simeq 0$.
Provided that the OP of NCCO has an $s$-wave symmetry, a strong reduction of
the density of states below $V_g$ is expected going to low $T$ resulting in an
increase of the $dV/dI$ at small $V$.  We note that an almost $T$ independent
$R_n$ has been observed for most GBJs fabricated from the hole doped
HTS\cite{Gross:95,Froehlich:97a}.  This difference may be related to a
different OP symmetry in the electron and hole doped HTS.  It is likely that
for the latter a $d$-wave symmetry of the OP together with pair breaking
effects at surfaces of $d$-wave superconductors cause a large and weakly $T$
dependent density of states around the Fermi level and, hence, a weakly $T$
dependent $R_n$.


\vspace*{-0.2cm}
\begin{figure}[b]
\noindent\hspace*{0cm}
\centering\epsfxsize=9cm\epsffile{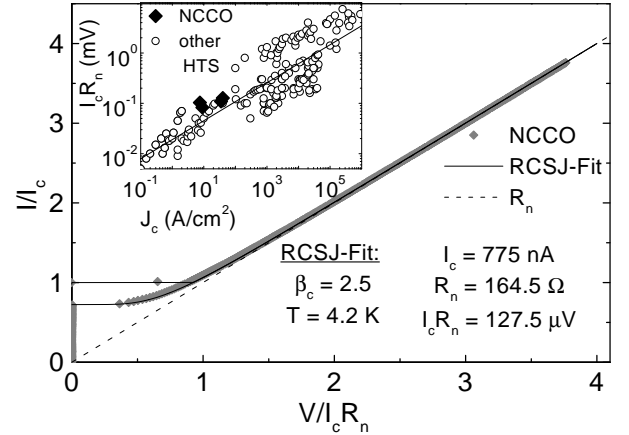}\vspace*{-0.3cm}
\caption[]{\small Current vs.  voltage of a $10\,\mu$m wide, 24$^{\circ}$ [001] tilt
NCCO bicrystal GBJ.  Also shown is the RCSJ-model prediction (solid line) and
the ohmic line (broken line).  In the inset, the $I_cR_n$ products of
NCCO-GBJs (full symbols) are plotted vs.~$J_c$ together with $I_cR_n$ products
of YBCO- and BSCCO-GBJs (bicrystal, step-edge, biepitaxial) taken from
Ref.~\cite{Gross:95} (open symbols).  }
\label{fig:iu}
\vspace*{-0.4cm}
\end{figure}


A typical IVC of a NCCO-GBJ is shown in Fig.~\ref{fig:iu}.  Due to the small
$J_c$ and large $\rho _n$ values of NCCO-GBJs, for a typical junction size of
a few $\mu$m$^2$ typical $I_c$ and $R_n$ values are below $1\,\mu$A and above
$100\,\Omega$, respectively, at 4.2\,K resulting in $I_cR_n$ products around
$100\,\mu$V small compared to $V_g$.  At 4.2\,K the IVC is slightly
hysteretic. Within the resistively and capacitively shunted junction (RCSJ)
model \cite{Likharev:86} a good fit is obtained for a McCumber-parameter
$\beta_C \simeq 2.5$.  We also note that the IVCs of NCCO-GBJs show no excess
current.  The inset of Fig.~\ref{fig:iu} clearly shows that the NCCO-GBJs fit
well to the general scaling relation $I_cR_n
\propto (J_c)^{q}$ with $q
\approx0.5\pm0.1$ found for hole doped HTS \cite{Gross:95}.  This strongly
suggests that the mechanism of charge transport across the grain boundary
barrier is similar for the hole and electron doped materials.  Furthermore,
since NCCO likely has a dominating $s$-wave and the hole doped HTS a
dominating $d$-wave component of the OP, the symmetry of the OP can be ruled
out as a main cause for the small value and the scaling relation of
the $I_cR_n$ product. A similar argument holds for the strong decrease of
$J_c$ with increasing misorientation angle. The similarity of the transport
properties of GBJs fabricated from the electron and hole doped HTS can be
naturally explained within the ISJ-model \cite{Gross:95,Gross:91a,Marx:97}.
Here, the small $I_cR_n$ values and the scaling behavior is explained by the
fact that for both NCCO-GBJs and those fabricated from the hole doped HTS the
main transport mechanism is resonant tunneling via localized states within an
insulating grain boundary barrier for the quasiparticles and direct tunneling
for the Cooper pairs.


\begin{figure}[tb]
\vspace*{-0.2cm}
\noindent\hspace*{0cm}
\centering\epsfxsize=9cm\epsffile{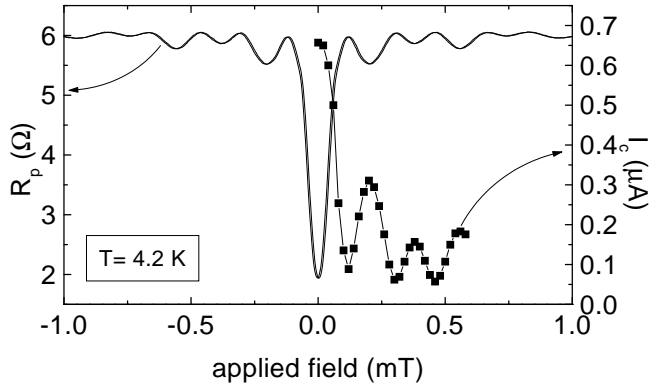}\vspace*{-1.0cm}
\caption[]{\small Magnetic field dependence of the critical current
of a $24^{\circ}$ [001] tilt NCCO-GBJ at 4.2\,K.
Solid symbols: $I_c$ values directly obtained from IVCs. Solid line: $I_c(B)$
curve derived from $R_p(B)$.
}
\label{fig:icb}
\vspace*{-0.25cm}
\end{figure}


Fig.~\ref{fig:icb} shows the magnetic field dependence of the critical
current.  The solid symbols represent the $I_c(B)$ data obtained by
determining $I_c$ directly from the IVCs.  The solid line shows $R_p(B)$.
Since at constant temperature $R_p(B)=R_n
\left[{\rm I_0}\left(\frac{\hbar I_c(B)}{2ek_BT}\right)
\right]^{-2}$ is determined only by $I_c(B)$, the field dependence of $I_c$
can be derived from the measured $R_p(B)$ dependence by inverting the Bessel
function \cite{Schuster:93}. The quantitative agreement is good and
Fig.~\ref{fig:icb} shows that both methods give the same modulation period as
expected \cite{Schuster:93}. The observed $I_c(B)$ pattern is perfectly
symmetric about $B=0$ and close to a Fraunhofer diffraction pattern expected
for small Josephson junctions with spatially homogeneous $J_c$.  This suggests
that $J_c$ is homogeneous along the grain boundary on a $\mu$m scale.

In conclusion, the superconducting properties of symmetric [001] tilt GBJs
fabricated from both the hole and electron doped HTS are very similar and can
be well described within the ISJ-model.  In particular, the scaling law
$I_cR_n \propto \sqrt{J_c}$ as well as the strong decrease of $J_c$ with
increasing misorientation angle found for the hole doped HTS also holds for
NCCO-GBJs.  Provided that the OP symmetry of electron and hole doped HTS is
$s$ and $d$-wave, respectively, this suggests that the grain boundary
properties are not strongly influenced by the symmetry of the OP but are
dominated by the structural properties of the grain boundary barrier that may
be similar for both materials.

The authors acknowledge valuable discussion with J. Halb\-rit\-ter. This work
is supported by the Deutsche For\-schungs\-ge\-mein\-schaft (SFB 341).

\end{multicols}


\vspace*{-0.75cm}
\begin{thebibliography}{99}
\vspace*{-2cm}
\small

\bibitem{Dimos:90}
D.~Dimos, P.~Chaudhari, and J.~Mannhart, Phys.~Rev.~{\bf B 41}, 4038 (1990).

\bibitem{Gross:95}
R.~Gross, in {\em Interfaces in Superconducting Systems}, S.~L.~Shinde and
D.~Rudman eds, Springer-Verlag, New York (1994), pp.~176-209; see also R.
Gross et al., IEEE Trans.~Appl.~Supercond.~{\bf 7}, 2929 (1997).

\bibitem{Beck:96}
A.~Beck, O.~M.~Fr\"{o}hlich, D.~K\"{o}lle, R.~Gross, H.~Sato, and M.~Naito,
Appl.~Phys.~Lett.~{\bf 68}, 3341 (1996).

\bibitem{Kawasaki:93}
M.~Kawasaki, E.~Sarnelli, P.~Chaudhari, A.~Gupta, A.~Kussmaul, J.~Lacey, and
W.~Lee, Appl. Phys. Lett. {\bf 62}, 417 (1993).

\bibitem{Gross:91a}
R.~Gross and B.~Mayer, Physica {\bf C 180}, 235 (1991).

\bibitem{Marx:97}
A.~Marx, L.~Alff, and R.~Gross, IEEE Trans.~Appl.~Supercond.~{\bf 7}, 2719
(1997).

\bibitem{Gross:90a}
R.~Gross, P.~Chaudhari, M.~Kawasaki, and A.~Gupta, Phys.~Rev.~{\bf B 42},
10735 (1990).

\bibitem{vHarlingen:95}
D.~J.~Van Harlingen, Rev.~Mod.~Phys.~{\bf 67}, 515 (1995).

\bibitem{Mannhart:96}
J.~Mannhart, H.~Hilgenkamp, B.~Mayer, C.~Gerber, J.~R.~Kirtley, K.~A.~Moler,
and M.~Sigrist, Phys.~Rev.~Lett.~{\bf 77}, 2782 (1996); see also Phys.~Rev.
{\bf B 53}, 14 586 (1996).

\bibitem{Huang:90}
Q. Huang, J. F. Zasadzinski, N. Tralshawala, K. E. Gray, D. G. Hinks, J. L.
Peng, and R. L. Greene, Nature {\bf 347}, 369 (1990).

\bibitem{Wu:93}
D. H. Wu, J. Mao, J. L. Peng, X. X. Xi, T. Venkatesan, R. L. Greene, and S. M.
Anlage, Phys. Rev. Lett. {\bf 70}, 85 (1993).

\bibitem{Andreone:94}
A. Andreone, A. Cassinese, A. Di Chiara, R. Vaglio, A. Gupta, and E. Sarnelli,
Phys. Rev. {\bf B 49}, 6392 (1994).

\bibitem{Alff:97}
L.~Alff, A.~Beck, A.~Marx, S.~Kleefisch, T.~Bauch, H.~Sato, M.~Naito,
G.~Koren, and R.~Gross (unpublished).

\bibitem{Naito:95}
M.~Naito and H.~Sato, Appl.~Phys.~Lett.~{\bf 67}, 2557 (1995).

\bibitem{Yamamoto:97}
H.~Yamamoto, M.~Naito, and H.~Sato, Phys.~Rev.~{\bf B 56}, 2852 (1997).

\bibitem{Ambegaokar:69}
V.~Ambegaokar and B.~I.~Halperin, Phys.~Rev.~Lett.~{\bf 22}, 1364 (1969).

\bibitem{Gross:90}
R.~Gross, P.~Chaudhari, D.~Dimos, A.~Gupta, and G.~Koren,
Phys.~Rev.~Lett.~{\bf 64}, 228 (1990).

\bibitem{Gapvoltage}
presuming a highly anisotropic $s$-wave OP for NCCO,
$V_g\simeq \Delta _0/e$ is expected instead of $V_g \simeq 2\Delta _0/e$.

\bibitem{Alff:96c}
L.~Alff, H.~Takashima, S.~Kashiwaya, N.~Terada, T.~Ito, K.~Oka, M.~Koyanagi,
and Y.~Tanaka, in {\em Advances in Superconductivity IX}, S.~Nakajima and
M.~Murakami eds, Springer-Verlag, Tokyo (1997), p.~49.

\bibitem{Naito:98}
M.~Naito, H.~Sato, and H.~Yamamoto, Physica {\bf C 293}, 36 (1997).

\bibitem{Froehlich:97a}
O.~M.~Fr\"{o}hlich, P.~Richter, A.Beck, R.~Gross, and G.~Koren,
J.~Low Temp.~Phys.~{\bf 106}, 243 (1997);
see also IEEE Trans. Appl. Supercond. {\bf 7}, 3189 (1997).

\bibitem{Likharev:86}
K.~K.~Likharev, {\em Dynamics of Josephson Junctions and Circuits}, Gordon and
Breach, New York (1986).

\bibitem{Schuster:93}
S.~Schuster, R.~Gross, B.~Mayer, and R.~P.~Huebener, Phys.~Rev.~{\bf B 48},
16172 (1993).

\end{thebibliography}
\end{document}